**Ion-Engineered Insulator-to-Semiconductor Transition in Natural 2D Biotite**

*Dipanwita Mitra, Raphael B. de Oliveira, Guilherme S. L. Fabris, Debkanta Ghosh, AyonJyoti Karmakar, Raphael M. Tromer, Marcelo L. Pereira Junior, Douglas S. Galvão*, Chandra Sekhar Tiwary*, Prasanta Kumar Datta**

Dipanwita Mitra, Debkanta Ghosh, AyonJyoti Karmakar, Prasanta Kumar Datta

Department of Physics

Indian Institute of Technology Kharagpur

Kharagpur, 721302, India

E-mail: pkdatta@phy.iitkgp.ac.in

Guilherme S. L. Fabris, Douglas S. Galvao

Department of Applied Physics and Center for Computational Engineering and Sciences

State University of Campinas

Campinas, São Paulo, Brazil

E-mail: galvao@ifi.unicamp.br

Raphael B. de Oliveira,

Department of Materials Science and Nano Engineering

Rice University, Houston, Texas, USA

Raphael M. Tromer

Institute of Physics,

University of Brasília, Brasília, DF, Brazil

Marcelo L. Pereira Junior

Department of Materials Science and Nano Engineering

Rice University, Houston, Texas, USA

College of Technology, University of Brasília, Brasília, DF, Brazil

Chandra Sekhar Tiwary

Department of Metallurgical and Materials Engineering

Indian Institute of Technology Kharagpur

Kharagpur, 721302, India

E-mail: chandra.tiwary@metal.iitkgp.ac.in

**Abstract Text**

Naturally occurring layered silicates offer an abundant yet unexplored class of 2D materials, but their insulating nature limits their functional utility. Here, we demonstrate a chemical strategy that transforms liquid-phase-exfoliated biotite nanosheets into a tunable 2D semiconductor through controlled NaOH treatment. The resulting insulator-to-semiconductor transition originates from Na incorporation, defect generation, and local structural reconstruction while largely preserving the layered framework. Structural and chemical analyses reveal lattice distortion, interlayer reorganization, hydroxylation, and partial $Na^{+}$–$K^{+}$ exchange, establishing the origin of the electronic restructuring. This transformation broadens the optical response, shifting the ~221 nm absorption toward ~280 and ~975 nm, reducing the optical bandgap from ~5.2 to ~3.2–3.5 eV, and introducing low-energy transitions at ~1.12–1.17 eV. Electrical measurements reveal nonlinear transport with currents reaching ~10 μA, demonstrating activated carrier conduction. Ultrafast transient absorption reveals pronounced excited-state absorption, with carrier cooling (0.16–0.38 ps) followed by fast (35–60 ps) and long-lived (336–491 ps) relaxation associated with trap-mediated recombination. Fluence-dependent dynamics reveal a hot-phonon bottleneck at elevated carrier densities. Together with density functional theory calculations, these results establish chemical defect and ion engineering as a powerful route for converting naturally abundant layered minerals into electronically tunable 2D materials for emerging optoelectronic and ultrafast photonic technologies.

## Introduction

Two-dimensional (2D) materials have significantly advanced the landscape of condensed matter and materials science by enabling unprecedented control over electronic, optical, and interfacial phenomena. Since the experimental realization of graphene[1], the field has expanded to include a variety of different families of layered systems, such as Transition Metal Dichalcogenides (TMDs)[2], hexagonal boron nitride[3], black phosphorus [4,5], MXenes[6], etc. These materials exhibit properties fundamentally distinct from their 3D bulk counterparts, unlocking broad opportunities in nanoelectronics[2], optoelectronics[2,5], energy storage[7], catalysis[8,9], and chemical sensing[10]. Their reduced dimensionality and strong light–matter interactions position them as key building blocks for next-generation photonic and optoelectronic technologies.

Beyond conventional synthetic 2D materials such as graphene and TMDs, naturally occurring silicate minerals have emerged as a structurally distinct and compositionally versatile class of low-dimensional systems. Unlike van der Waals solids, these materials encompass both layered and non-layered silicate systems, thereby extending the accessible structural and chemical space for property engineering. This intrinsic heterogeneity enables emergent functionalities that are not readily achievable in conventional 2D systems, offering a platform to explore structure–property relationships in chemically and structurally complex environments. In this context, 2D silicate minerals are increasingly recognized as promising candidates for electronic[11], optoelectronic[12,13], and energy-related applications[14], demonstrating tunable electronic and optical functionalities governed by defect and interfacial chemistry at reduced dimensionality.

In our recent work, few-layer biotite was shown to exhibit an exceptionally strong nonlinear optical response[15], with a two-photon absorption coefficient of $(9.75 \pm 0.15) \times 10^5$ cm $GW^{-1}$ at 415 nm under femtosecond excitation and an optical limiting threshold of 1.51 mJ $cm^{-2}$, surpassing benchmark 2D materials, including graphene and transition metal dichalcogenides. Comparable nonlinear optical and optical limiting responses have also been observed across other silicate systems, including muscovite[12] (layered) and rhodonite[13] (non-layered), highlighting the broader potential of naturally occurring silicate minerals for nonlinear photonic functionalities.

Structurally, biotite consists of tetrahedral silicate sheets sandwiching Fe/Mg-containing octahedral layers, stabilized by interlayer $K^+$ ions[15]. This architecture enables exfoliation into atomically thin nanosheets via liquid-phase methods, providing direct access to low-dimensional mineral systems[15]. However, despite this structural accessibility, pristine biotite remains a wide-bandgap insulator (~5.0 eV) with negligible free-carrier density, which fundamentally limits its applicability in electronic and optoelectronic systems. This sharp contrast between structural versatility and electronic inertness highlights the necessity of strategies capable of driving an insulator-to-semiconductor transition, thereby enabling active carrier transport.

In layered silicates, defect states and interlayer ionic species[16] play a central role in governing charge localization, recombination, and transport. When appropriately engineered, these features provide a pathway to reconstruct the electronic structure, enabling a controlled insulator-to-semiconductor transition accompanied by emergent carrier dynamics. In this context, coupled defect engineering and chemical doping offer an effective route to activate charge transport[17] in otherwise electronically inert mineral frameworks[18].

Among the various approaches, alkali treatment provides a simple and scalable chemical strategy for such modification[18,19]. In silicate frameworks, NaOH facilitates ion exchange between interlayer $K^+$ and $Na^+$, induces local lattice distortion, and modifies the electrostatic environment. These coupled processes generate defect states and redistribute charge density, thereby reconstructing the electronic structure and enhancing optically driven carrier responses.

Here, we demonstrate controlled electronic modulation in liquid-phase exfoliated biotite nanosheets through NaOH treatment. A comprehensive suite of structural, chemical, and spectroscopic techniques, including X-ray diffraction (XRD), Raman spectroscopy, X-ray photoelectron spectroscopy (XPS), high resolution transmission electron microscopy (HRTEM), scanning transmission electron microscopy with energy-dispersive spectroscopy (STEM–EDS), UV–visible spectroscopy, electrical transport (I–V) measurements, ultrafast transient absorption spectroscopy, and density functional theory (DFT) calculations, provide a unified understanding of alkali-induced electronic reconstruction. We show that the synergistic interplay between defect formation and Na incorporation drives a pronounced insulator-to-semiconductor transition, leading to the emergence of semiconducting behavior and significantly altering carrier dynamics. Collectively, these findings establish a general chemical strategy for engineering the electronic properties of naturally abundant silicate minerals. More broadly, this work positions earth-abundant mineral systems as a chemically tunable and structurally diverse platform for sustainable optoelectronic and photonic functionalities.

## Results and Discussion

**Figure 1(a)** presents the XRD patterns of NaOH, pristine 2D biotite, and NaOH-treated 2D biotite, clearly revealing structural modification induced by the chemical treatment. NaOH crystallizes in an orthorhombic structure (JCPDS No. 96-231-0701) with lattice parameters: $a$ = 6.05 Å, $b$ = 11.72 Å, and $c$ = 6.21 Å, and interaxial angles of α = β = γ = 90°. Pristine 2D biotite exhibits a monoclinic crystal structure (JCPDS No. 96-900-1353) with space group C12/m1, characterized by lattice parameters: $a$ = 5.33 Å, $b$ = 9.25 Å, and $c$ = 10.18 Å, with α = γ = 90° and β = 100.17°. Upon treatment with 5 M NaOH, the 2D biotite retains its monoclinic symmetry (JCPDS No. 96-900-1351) while exhibiting slight lattice contraction, with redefined lattice parameters: $a$ = 5.31 Å, $b$ = 9.19 Å, and $c$ = 10.16 Å, and α = γ = 90°, β = 100.18°. The decrease in lattice parameters indicates modification of the interlayer environment induced by the NaOH treatment.

A detailed view of the (003) reflection in **Figure 1(b)** shows that the single diffraction peak observed for pristine 2D biotite evolves into multiple components in the treated samples. The (003) reflection of 2D biotite is centered at ~26.4°, characteristic of its interlayer spacing. Upon NaOH treatment, the peak evolves into an asymmetric shape that can be deconvoluted into three components located at ~26.36°, 26.46°, and 26.6°, revealing the emergence of heterogeneous interlayer configurations[20,21]. In accordance with Bragg's law $\left(2d\sin\theta = n\lambda\right)$, the shift toward higher diffraction angles signifies a reduction in interlayer spacing. Specifically, the lower-angle component (~26.36°) corresponds to locally expanded regions, whereas the higher-angle components (~26.46° and 26.6°) indicate progressively contracted domains. The pronounced redistribution of intensity toward higher 2θ values indicates a net lattice contraction along the stacking direction. This structural reorganization is attributed to $Na^+$-mediated modulation of interlayer interactions, accompanied by partial removal and rearrangement of intercalated species[21,22], resulting in a more compact and electronically altered layered framework.

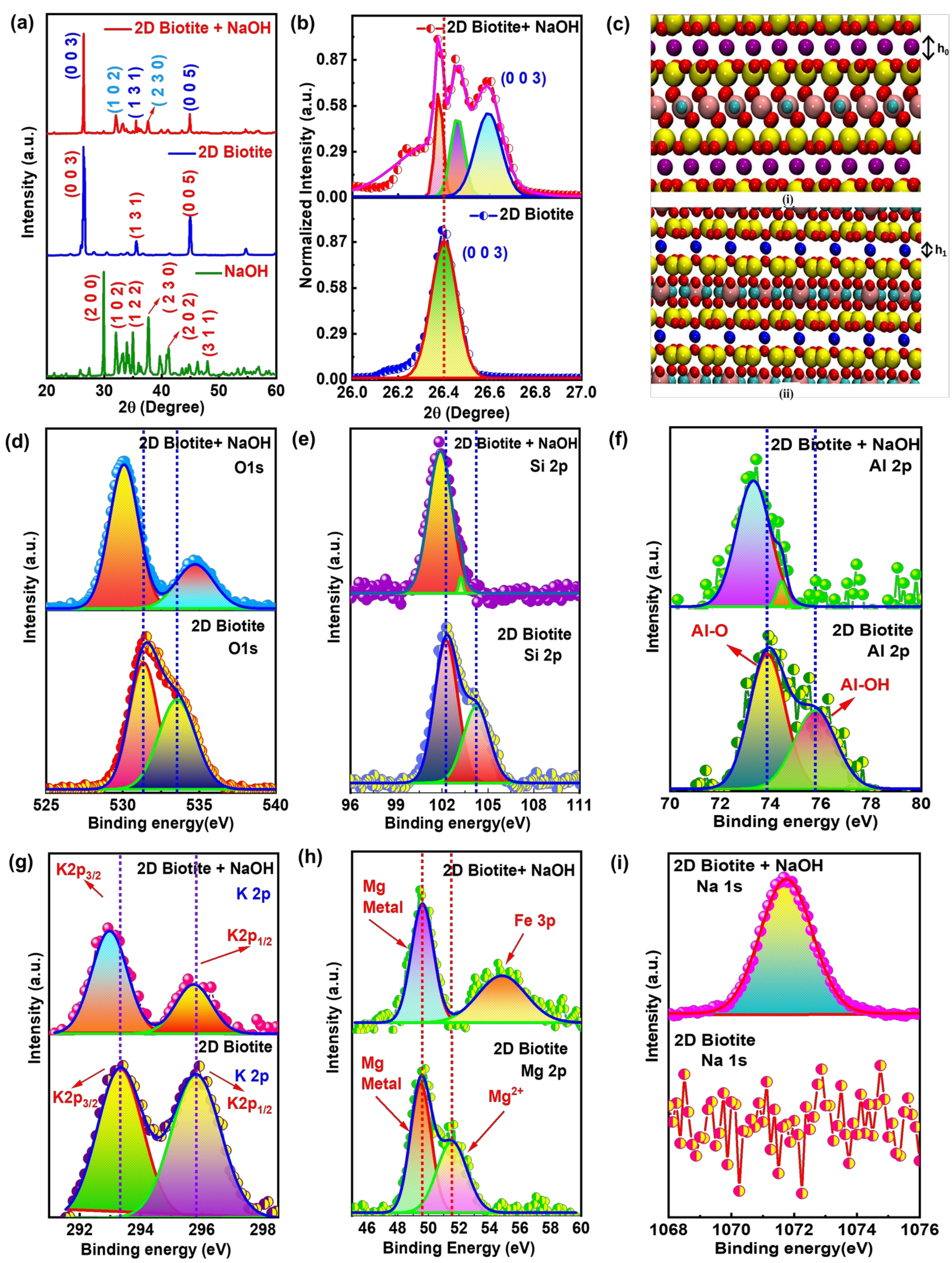

***Figure 1.*** *(a) XRD patterns of NaOH, pristine 2D biotite, and NaOH-treated 2D biotite;(b) Comparison of the (003) XRD peak of 2D biotite and NaOH-treated 2D biotite, showing a single peak for pristine biotite and splitting into three components after treatment. (c) Optimized structures of standard and Na-modified biotite after geometry relaxation. Atomic species are represented as silicon (yellow), oxygen (red), potassium (purple), magnesium (pink), iron (cyan), and sodium (blue), respectively. XPS spectra comparing pristine 2D biotite and NaOH-treated 2D biotite for (a) O 1s, (b) Si 2p, (c) Al 2p, (d) K 2p, (e) Mg 2p, and (f) Na 1s core levels. The NaOH treatment induces clear changes in binding energies, peak intensities, and chemical states, indicating surface chemical modification, partial ion exchange ($K^+ \leftrightarrow Na^+$), and altered metal–oxygen bonding environments in the treated 2D biotite.*

The lattice contraction observed in XRD is further corroborated by DFT calculations, supporting the proposed structural reorganization. The initial stage of the analysis focused on examining the structural changes induced by sodium incorporation in biotite. Two configurations were considered to capture the effects of this modification. The first corresponds to the standard biotite unit cell, in which potassium (K) atoms occupy the interlayer sites and serve as a structural reference. The second configuration, referred to as modified biotite, was obtained by replacing the interlayer potassium atoms with sodium (Na) atoms to evaluate the structural rearrangements induced by this substitution.

**Figure 1(c)** shows the optimized structures obtained after complete structural relaxation. The interlayer distance between two consecutive van der Waals planes was used as a measure of the structural changes. For the standard biotite, the interlayer separation ($h_0$) was estimated to be 4.45 Å, whereas for the Na-modified configuration, it decreased to 4.16 Å, indicating a contraction of approximately 6.5%. These distances are highlighted in **Figures 1(c)(i)** and **1(c)(ii)**, represented as $h_0$ and $h_1$, respectively. This contraction can be attributed mainly to the difference in the van der Waals radii of the cations. Potassium, with a radius of 2.75 Å, is considerably larger than sodium's, which is 2.27 Å, representing a reduction of about 17%. The smaller ionic radius of sodium allows the layers to come closer to each other, resulting in a more compact equilibrium configuration with a reduced interlayer spacing.

To validate the theoretical model, the optimized lattice parameters were compared with experimental data and previously reported values, as summarized in **Table S1**. For pristine biotite, the optimized lattice parameters exhibited deviations of 0.37%, 0.43%, and 0.29% for $a$, $b$, and $c$, respectively, with an angular deviation of 0.46% in $\beta$. For the Na-modified structure, the deviations were 1.51%, 0.65%, and 2.85% for the lattice parameters, with a $\beta$ deviation of 1.10%. These small deviations demonstrate the reliability of the DFT model in reproducing the structural characteristics of biotite and its Na-modified counterpart.

Since the experimental treatment involves exposure of exfoliated biotite to NaOH solution, the theoretical model focused on the interaction between Na species and the exposed biotite surface. The adsorption energy landscape shown in **Figure S1(a)** reveals periodic adsorption minima located at the intrinsic pore sites of the 2D biotite lattice. These energetically favorable sites provide strong electrostatic stabilization for Na incorporation through interaction with surrounding oxygen atoms. The optimized adsorption geometry presented in **Figure S1(b)** confirms that Na atoms preferentially occupy these pore regions without inducing significant structural distortion. The calculated Na–O bond distances range from 2.2 Å to 2.3 Å, consistent with stable ionic coordination within the layered framework.

To investigate the effect of Na concentration, multiple adsorption configurations containing increasing numbers of Na atoms were considered, corresponding to Na concentrations ranging from 2.8% to 16.8% (**Figures S2(a–f)**). Successive Na incorporation was modelled by progressively occupying energetically favorable pore sites followed by structural relaxation. The optimized structures reveal that Na atoms can be accommodated within the interlayer cavities up to approximately 11.2% concentration without destabilizing the lattice. Beyond this concentration, additional Na atoms preferentially remain adsorbed on the surface after relaxation, indicating a transition from interlayer incorporation to surface adsorption. Detailed concentration calculations, structural configurations, and optimized geometries are discussed in the Supplementary Information. The thickness of the exfoliated 2D biotite nanosheets was characterized by AFM, with corresponding images provided in the Supporting Information (**Figure S3**).

**Figure 1(d)** presents the deconvoluted O1s spectrum of pristine 2D biotite, showing two components centered at ~531.35 eV and ~533.59 eV, corresponding to lattice oxygen (Si–O–Si/Al–O) and surface hydroxyl or adsorbed species, respectively. Upon NaOH treatment, the lattice oxygen peak shifts to lower binding energy (~530 eV), indicating an increase in electron density due to Na–O interaction and possible formation of Na-modified silicate environments[23]**.** Concurrently, the high binding energy component shifts to ~534.72 eV, suggesting enhanced hydroxylation and the presence of strongly bound $OH^-$ or hydrated species[24]. These results confirm significant surface chemical modification and ion-induced restructuring of the 2D biotite layers.

**Figure 1(e)** illustrates that the Si 2p spectrum of pristine 2D biotite exhibits two components at approximately 102.32 eV and 104.24 eV. The peak at ~102.32 eV is attributed to $Si^{4+}$ in the aluminosilicate framework, corresponding to Si–O–Si and Si–O–Al bonding environments[25]**,** while the higher binding energy component at ~104.27 eV arises from more oxidized $Si^{4+}$ species, such as surface Si–$O_x$ or Si–OH groups. Upon NaOH treatment the Si 2p peaks shift to lower binding energies, appearing at ~101 eV and ~103.17 eV. The dominant peak at ~101 eV indicates that most of the Si atoms reside in an electron-rich environment, consistent with the formation of Si–$O^-$ species stabilized by $Na^+$ ions. In contrast, the higher binding energy component at ~103.17 eV is significantly reduced in intensity, suggesting that only a minor fraction of the original aluminosilicate framework remains unmodified. These results indicate substantial electronic and structural modification of the silicate network induced by the NaOH treatment.

**Figure 1(f)** demonstrates that the K 2p spectrum of pristine 2D biotite exhibits two peaks at approximately 293.32 eV and 295.81 eV, corresponding to the spin–orbit-split K $2p_{3/2}$ and K $2p_{1/2}$ components of $K^+$ ions present in the interlayer sites of the biotite structure. Upon NaOH treatment, these peaks shift slightly to lower binding energies, appearing at ~292.95 eV and ~295.74 eV. This small shift indicates a marginal increase in electron density around the potassium ions, likely arising from changes in the local chemical environment due to $Na^+$ incorporation and partial ion exchange or structural modification of the interlayer region. The reduced intensity of the K 2p XPS component in the NaOH-treated 2D biotite indicates partial depletion of interlayer $K^+$ ions, likely due to $Na^+$–$K^+$ ion exchange[26] and alkaline-induced leaching, consistent with structural exfoliation and surface modification.

**Figure 1(g)** presents the deconvoluted Al 2p spectrum, with the main peak centered at approximately 73.8 eV, attributed to $Al^{3+}$ species associated with Al–O (Al–O–Si) bonds in the aluminosilicate lattice, consistent with reported values for $Al_2O_3$ (~74.1 eV) [27]and aluminosilicates (~74.4 eV)[28]. The higher binding energy component at ~75.8 eV is assigned to surface hydroxylated Al species (Al–OH), in agreement with reported Al 2p positions in AlO(OH)[29].

Upon NaOH treatment, the Al 2p spectrum exhibits a shift of the main peak to ~73.32 eV, accompanied by a minor component at ~74.4 eV. The negative shift in binding energy suggests a modification of the local chemical environment of Al, likely arising from interaction with $Na^+$ ions, increased electron density around Al centers, and structural distortion of the aluminosilicate framework. The component at ~74.4 eV corresponds to lattice $Al^{3+}$ species (Al–O–Si), indicating partial retention of the original framework. This shift, along with the redistribution of spectral components, confirms chemical modification and possible surface reconstruction induced by NaOH treatment.

The deconvoluted Mg 2p spectrum **(Figure 1(h))** shows a peak at ~49.5 eV attributed to $Mg^{2+}$ in Mg–O bonds within the biotite lattice, while the higher binding energy component at ~51.5 eV is assigned to surface hydroxylated Mg species, consistent with $Mg(OH)_2$-like environments. In the NaOH-treated biotite, the Mg 2p spectrum retains the peak at ~49.5 eV, attributed to lattice $Mg^{2+}$ (Mg–O). Notably, the higher binding energy component at ~51.5 eV, previously assigned to surface hydroxylated Mg species (Mg–OH), disappears after treatment, indicating removal or transformation of surface hydroxyl groups. Additionally, a new feature observed at ~54.7 eV is attributed to the Fe 3p core level, suggesting increased exposure or contribution of Fe species upon NaOH treatment.

**Figure 1(i)** reveals that while pristine 2D biotite shows no detectable Na 1s signal, the NaOH-treated sample exhibits a pronounced peak at ~1071 eV, confirming the incorporation of $Na^+$ ions via ion exchange or adsorption at oxygen sites, resulting in surface and structural modifications.

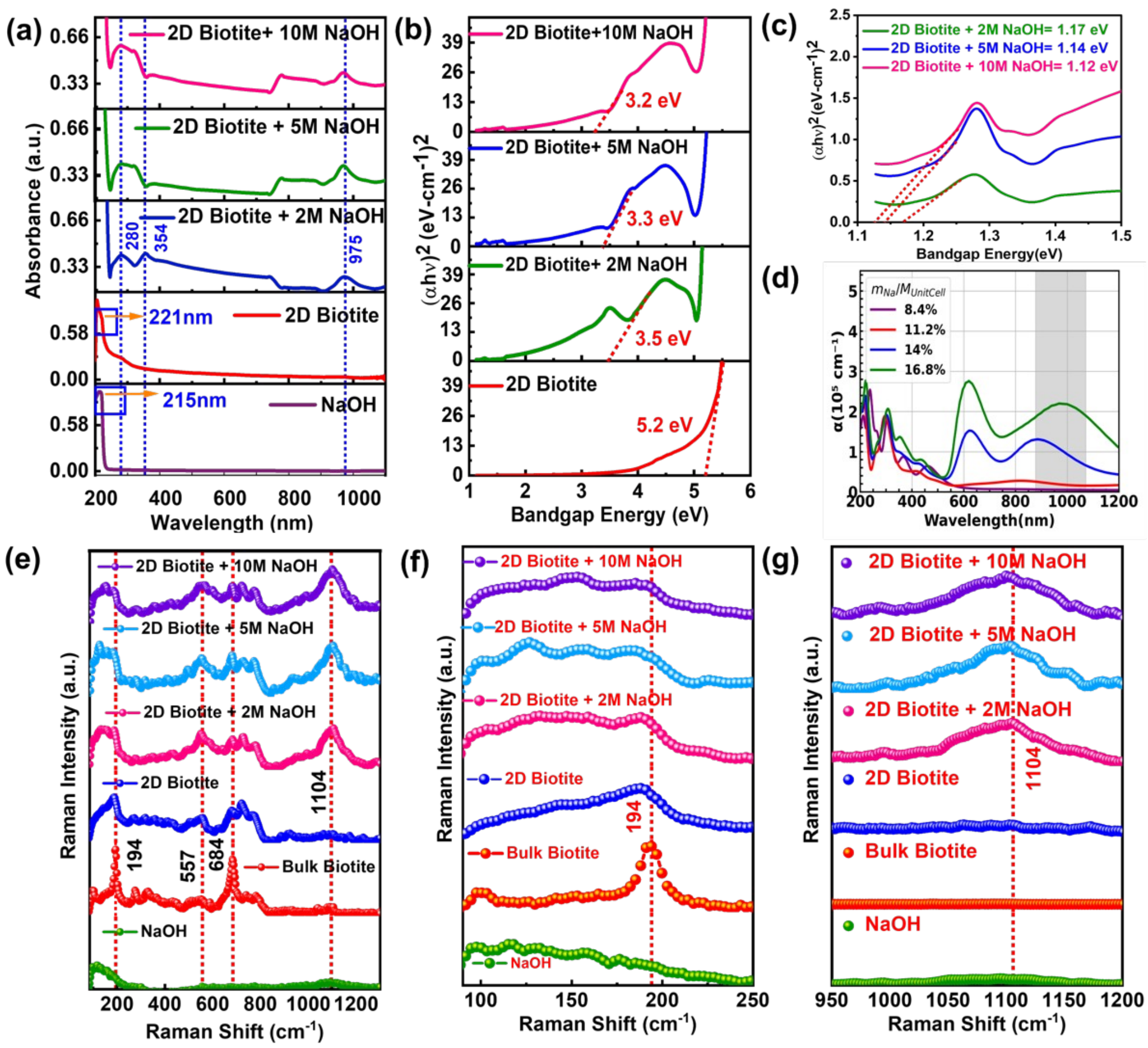

***Figure 2****. (a) UV–Vis spectra of 5M NaOH, 2D biotite and NaOH-treated biotite (2M, 5M, 10M), showing a pristine peak at ~221 nm and new absorption bands at ~280 nm and ~975 nm after treatment; (b) Tauc plots showing bandgap reduction from ~5.0 eV (pristine) to ~3.5, ~3.3, and ~3.2 eV after 2M, 5M, and 10M NaOH treatment, respectively;(c) Enlarged Tauc plots showing additional low-energy transitions at ~1.17–1.12 eV in treated samples; (e) Raman spectra of NaOH, bulk biotite, 2D biotite, and NaOH-treated 2D biotite; (d)Simulated optical absorption spectra for different Na concentrations in 2D biotite; (e) Raman spectra of NaOH, bulk biotite, 2D biotite, and NaOH-treated 2D biotite; (f) Enlarged Raman spectra of the ~194 $cm^{-1}$ mode showing redshift and broadening after exfoliation and NaOH treatment; (g) Raman spectra showing the emergence of a ~1104 $cm^{-1}$ mode after NaOH treatment, indicating chemical modification.*

***Figure 2(a)*** shows the UV–visible absorption spectra of pristine and NaOH-treated 2D biotite. A 5 M NaOH solution exhibits an absorption band at ~215 nm (enlarged version shown in ***Figure S4***), whereas pristine 2D biotite shows a characteristic absorption band at ~221 nm (enlarged version shown in ***Figure S5***), consistent with its wide-bandgap insulating nature. Upon treatment with 2 M NaOH, the ~221 nm feature disappears, and two distinct absorption peaks emerge at ~280 and ~354 nm, accompanied by a new near-infrared absorption band at ~975 nm. With increasing NaOH concentration to 5 and 10 M, the ~354 nm feature disappears, while the ~280 nm absorption band broadens toward longer wavelengths, extending to ~330 nm. This concentration-dependent spectral broadening and the disappearance of the distinct ~354 nm feature indicate increasing overlap and broadening of the electronic transitions associated with NaOH-induced modification of the biotite electronic structure. Furthermore, the emergence of the near-infrared absorption band at ~975 nm, which is absent in the NaOH solution control, suggests the formation of new sub-bandgap electronic states associated with NaOH-induced modification of the biotite lattice. These changes can be attributed to Na incorporation, accompanied by structural doping[30] and defect-induced electronic states, consistent with the sub-bandgap optical absorption reported in low-dimensional materials. **[31] [32]**.

In **Figure S6**, the DFT simulated absorption spectrum of pristine 2D biotite is shown. The simulated spectrum exhibits an absorption feature at approximately 298 nm, corresponding to an energy of 4.16 eV. This feature is red-shifted relative to the experimentally observed absorption feature at approximately 221 nm. The experimentally determined optical band gap of 5.20 eV, obtained from the Tauc plot, differs by approximately 20% from the DFT-calculated value of 4.16 eV. The underestimation is consistent with the known limitations of the GGA-PBE functional, which systematically underestimates the electronic bandgap values[33] . The pristine case serves as an essential reference for analyzing the changes introduced by Na incorporation.

The Tauc plots presented in **Figure 2(b)** further demonstrate a systematic reduction in optical bandgap from ~5.2 eV for pristine 2D biotite to ~3.5, ~3.3, and ~3.2 eV for samples treated with increasing NaOH concentrations, reflecting apparent bandgap narrowing arising from doping[30] [34]and defect-induced electronic states**[22,35]**. Such bandgap reduction, together with the emergence of low-energy absorption, suggests a transition of the material from an insulating to a semiconducting regime.[36] **Figure 2(c)** provides a magnified view of the Tauc plots for the treated samples, revealing additional low-energy transitions in the range of ~1.17

–1.12 eV, which correspond to infrared absorption features and further confirm the presence of localized states within the bandgap[31] [32,37]

**Figure 2(d)** shows the DFT simulated optical response for Na-doped and Na-adsorbed biotite at different concentrations. The region between 878 and 1073 nm, shaded in gray, corresponds to the experimental absorption band centered near 975 nm, allowing a qualitative comparison despite the inherent GGA-PBE underestimation. For Na concentrations below 8.4%, no significant absorption is observed in this spectral window. When the concentration exceeds 11.2%, optical activity appears within the shaded range, consistent with the previously identified saturation limit for Na incorporation. At this point, part of the Na is no longer structurally bonded and becomes adsorbed on the surface, generating additional electronic states near the Fermi level. These surface Na atoms contribute to the formation of new optical transitions, giving rise to the experimentally observed absorption band around 975 nm.

Raman spectra in **Figure 2(e)** provide detailed insight into the evolution of lattice dynamics from bulk to chemically modified biotite. The 5 M NaOH spectrum displays a broad low-frequency feature centered at ~115 $cm^{-1}$, arising from restricted translational dynamics of $OH^-$ within the hydrogen-bonded aqueous network, together with an O–H stretching mode at ~3279 $cm^{-1}$, highlighted in the magnified view in **Figure S7** [38]

Bulk biotite exhibits characteristic vibrational modes at ~194 $cm^{-1}$, attributed to internal vibrations of the $MO_6$ octahedra along with contributions from interlayer cation dynamics[39]. The mode at ~557 $cm^{-1}$ arises from the overlapping of (OH) and Si–O vibrations[39]. The prominent band at ~690 $cm^{-1}$, together with modes at ~741 and ~778 $cm^{-1}$, is assigned to Si–O–Si vibrations involving bridging oxygen atoms ($O_b$), which link adjacent $SiO_4$ tetrahedra and give rise to the layered framework of biotite[39,40].

Upon exfoliation to 2D biotite, the ~194 $cm^{-1}$ mode undergoes a red shift accompanied by spectral broadening , consistent with phonon confinement effects in the nanoscale regime (<10 nm). The associated relaxation of the $q \approx 0$ selection rule enables contributions from phonons away from the Brillouin zone center, resulting in asymmetric line shapes shifted toward lower wavenumbers. Within the phonon confinement model (PCM), these spectral evolutions directly correlate with reduced crystallite dimensions, providing a reliable probe of nanoscale structural modification in 2D biotite[41]. The enhanced intensity of the ~557 $cm^{-1}$ mode, accompanied by suppression of the ~684 $cm^{-1}$ mode in 2D biotite, reflects defect generation induced by exfoliation, leading to structural disorder and perturbation of the Si–O network[42].

With NaOH treatment (2 M, 5 M, and 10 M), these effects become more pronounced: the ~194 $cm^{-1}$ mode exhibits further red shift and significant broadening (**Figure 2(f)**), which can be attributed to NaOH-induced surface charge-transfer doping (SCTD) [43,44]**,** along with increased structural disorder, and enhanced phonon confinement.

The continued enhancement of the ~557 $cm^{-1}$ mode and attenuation of the ~684 $cm^{-1}$ band indicate progressive disruption of the silicate framework due to chemical doping/modification [45]. Notably, a new Raman band emerges at ~1104 $cm^{-1}$ **(Figure 2(g)),** which can be attributed to Na-induced lattice perturbation, where coupled doping and defect formation modify the Si–O bonding environment, break local symmetry[46], and activate additional vibrational modes. In addition, the appearance of a distinct O–H stretching mode at ~3420 $cm^{-1}$ **(Figure S7)**, blue-shifted relative to NaOH (~3279 $cm^{-1}$), signifies the formation of surface-bound hydroxyl groups with altered hydrogen-bonding environments. Such hydroxylation-induced vibrational shifts are consistent with changes in local bonding configuration and reduced hydrogen-bond strength at modified surfaces[47]. The DFT-simulated Raman spectra are provided in **Figure S8** of the Supplementary Information.

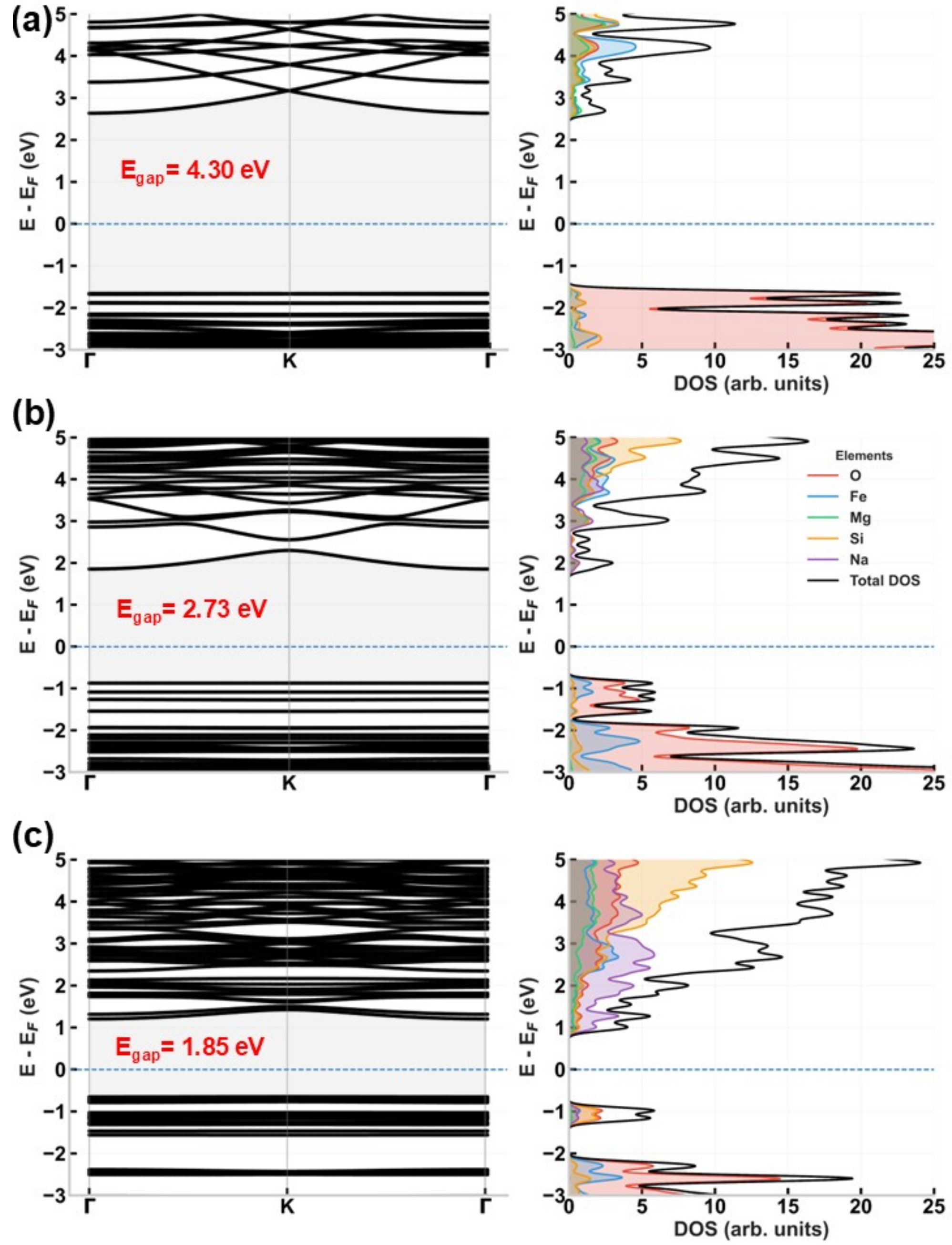

***Figure 3.*** *Electronic band structure and the corresponding projected density of states (PDOS) for (a) pristine 2D biotite, (b) 2.4% Na in 2D biotite, and (c) 8.4% Na in 2D biotite. Each Na insertion leads to a decrease of ~ 0.3 eV in the electronic bandgap.*

A similar behavior appears in electronic structure analysis. The band structure of pristine 2D biotite, shown in **Figure 3(a)**, confirms its insulating nature, exhibiting a direct bandgap of 4.30 eV. The valence band maximum is mainly dominated by O 2p states, while the conduction band minimum is largely composed of Fe 3d states, which together shape the total density of states (DOS). This distribution highlights the hybridization between Fe–O and Si–O bonds that stabilizes the layered biotite framework.

Upon Na incorporation, a progressive reduction in the electronic bandgap is observed. For a Na concentration of 2.4% (**Figure 3(b)**), the bandgap decreases from the pristine value of 4.30 eV to 2.73 eV, accompanied by the emergence of Na-derived states near the Fermi level, primarily associated with Na 3s orbitals. These states weakly affect the Fe–O electronic environment, leading to a small redistribution of charge between Na and the surface oxygen atoms.

At a higher Na concentration of 8.4% (**Figure 3(c)**), the electronic bandgap further decreases to 1.85 eV. The PDOS reveals that Na-related states become more pronounced near the conduction band edge, indicating stronger electronic coupling between the adsorbed Na atoms and the biotite lattice. This trend reflects an increase in electronic delocalization as Na loading increases.

Overall, the results indicate a systematic and monotonic decrease in the electronic bandgap as the Na concentration increases. With sufficient Na incorporation, the accumulation of Na-derived states near the Fermi level is expected to eventually close the gap entirely, driving the system toward a metallic behavior. This demonstrates that Na adsorption acts as an effective electronic dopant, progressively tuning the semiconducting 2D biotite toward a conductive regime.

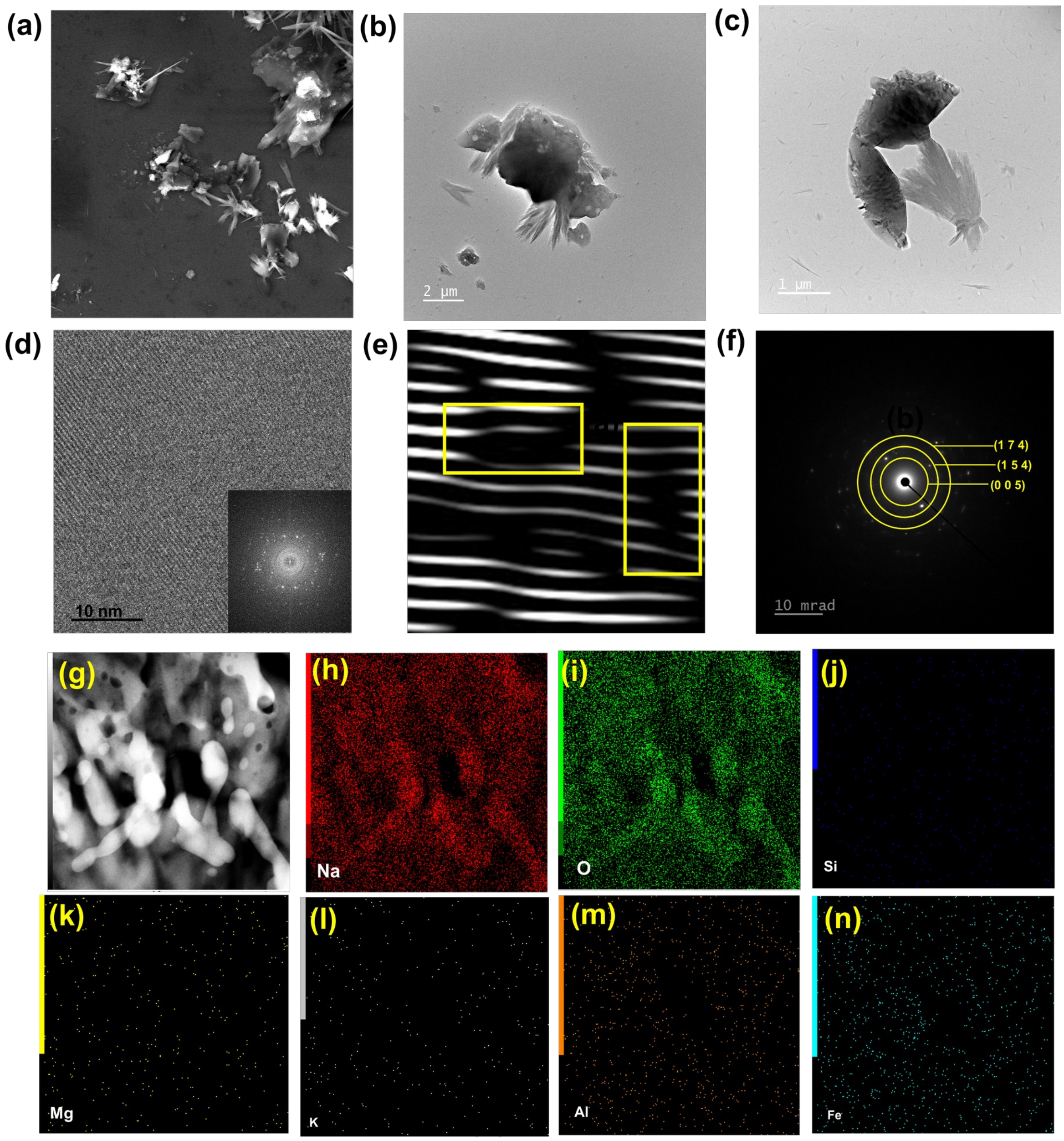

***Figure 4.*** *(a) SEM image of NaOH-treated 2D biotite; (b–c) low-magnification TEM images showing interaction between 2D biotite and NaOH; (d) HRTEM image of treated 2D biotite with inset FFT pattern; (e) inverse FFT image showing lattice defects and dislocations; and (f) SAED pattern of NaOH-treated biotite (g) STEM image of* NaOH-treated 2D biotite*; (h–n) STEM–EDS elemental maps of Na, O, Si, Mg, K, Al, and Fe, confirming Na incorporation and elemental redistribution.*

**Figure 4** illustrates the morphological and structural evolution of 2D biotite following NaOH treatment through complementary electron microscopy analyses. The SEM image **(Figure 4(a))** reveals noticeable changes in surface morphology, indicating alkali-induced modification of the layered structure. The low-magnification TEM image **(Figure 4(b))** shows clear interaction between the 2D biotite flakes and NaOH, suggesting chemical and structural reorganization at the flake surfaces. High-resolution TEM **(Figure 4(d))** resolves well-defined lattice fringes, confirming the retention of structural crystallinity after treatment. The corresponding FFT pattern **(Figure 4(d))** further supports the presence of ordered lattice planes, while the inverse FFT image **(Figure 4(e))** highlights a high density of lattice defects and dislocations generated by NaOH exposure. The SAED pattern (**Figure 4(f)**) exhibits concentric diffraction rings corresponding to different crystallographic planes, confirming the polycrystalline nature of NaOH-treated 2D biotite with multiple randomly oriented crystalline domains. A separate SAED pattern of the NaOH-treated 2D biotite (**Figure S9**) reveals a more diffuse nature of the diffraction rings, suggesting increased structural disorder and the presence of defects. Together, these results demonstrate that NaOH treatment induces defect formation and lattice distortion in 2D biotite while preserving its overall crystalline framework, which is crucial for the observed modification of its electronic properties.

**Figures 4(g-n)** present STEM–EDS analysis of the NaOH-treated 2D biotite, providing insight into its nanoscale elemental composition and distribution. The STEM image (**Figure 4g**) shows the overall morphology of the treated biotite flake, while the corresponding elemental maps **(Figure 4(g–n))** reveal the spatial distribution of Na, Al, O, Si, K, Mg, Fe, and F. The presence of a well-distributed Na signal confirms successful incorporation of Na species following NaOH treatment. Framework elements such as Si, Al, and O remain uniformly distributed, indicating preservation of the biotite lattice, whereas variations in the K signal suggest partial depletion of interlayer potassium. The homogeneous distribution of Mg, and Fe, and further supports the structural integrity of the octahedral and tetrahedral sites. Overall, the STEM–EDS results corroborate the XPS and TEM findings, demonstrating alkali-induced chemical modification of 2D biotite without phase segregation, which is essential for the observed changes in its electronic properties.

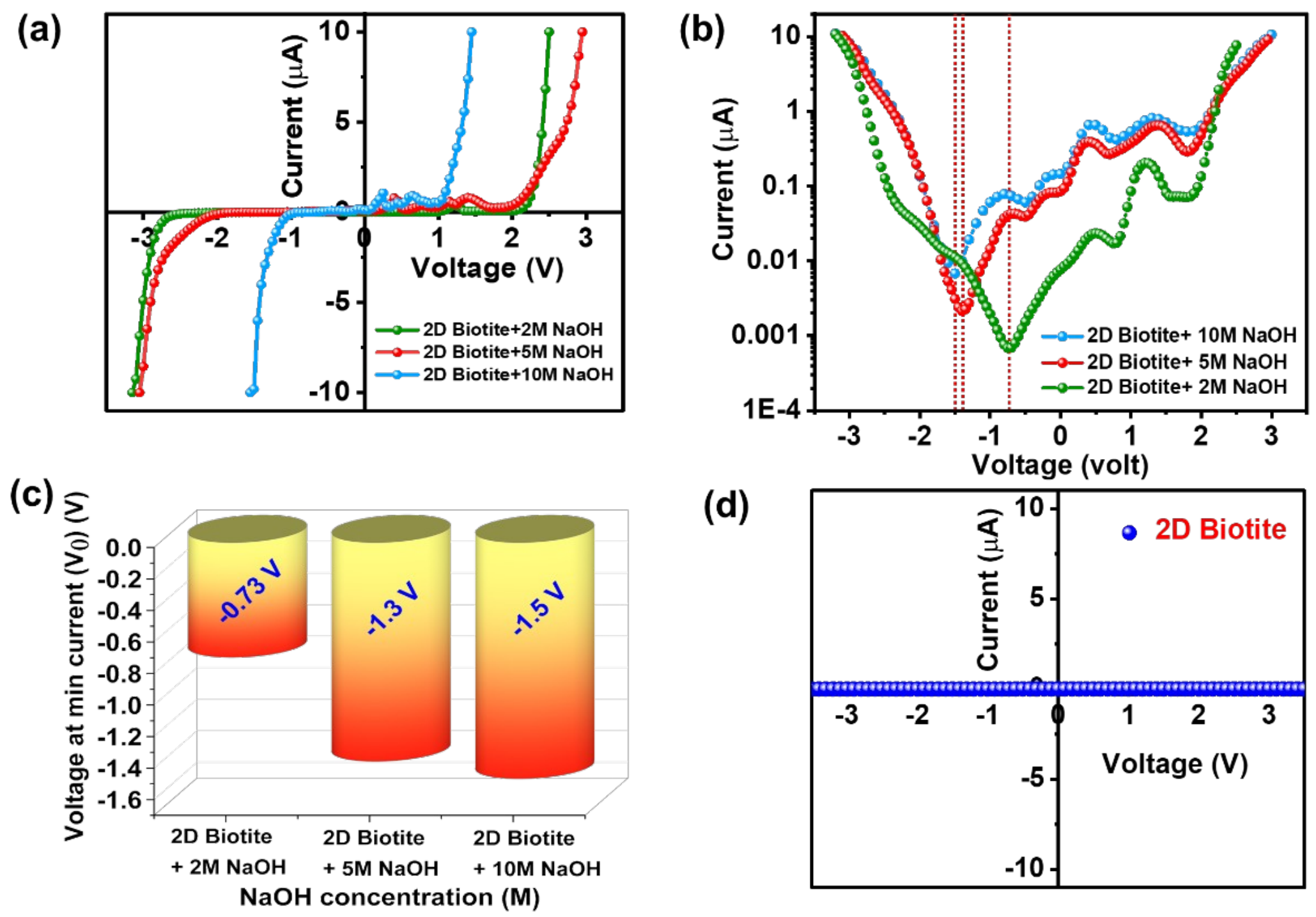

***Figure 5.*** *(a) Current–voltage (I–V) characteristics of 2D biotite treated with NaOH at different concentrations (2M, 5M, and 10M). (b) Corresponding semi-logarithmic plots (log|I| vs voltage) showing asymmetric behavior with the minimum current shifted from zero bias. (c) Extracted voltage at minimum current ($V_0$) as a function of NaOH concentration, showing a systematic shift toward negative bias with increasing concentration. (d) I–V characteristics of pristine 2D biotite for comparison.*

**Figure 5(a)** presents the I–V characteristics of NaOH-treated 2D biotite, revealing a pronounced threshold-type conduction behavior, in which the current remains close to zero within the µA range up to a critical bias, followed by an abrupt current rise. This non-linear response signifies the activation of field-assisted transport channels that are absent in pristine biotite. Notably, the threshold voltage decreases systematically with increasing NaOH concentration, indicating a progressive reduction in carrier injection barriers. This behavior is attributed to the synergistic effects of defect generation and $Na^+$ incorporation within the layered framework, which introduce electronically active states and facilitate charge transport, as corroborated by XPS and EDS analyses. Such chemically driven modulation of electronic structure parallels alkali-metal-induced conductivity enhancement in low-dimensional systems, where charge transfer and emergent electronic states fundamentally alter transport characteristics[48,49].

The semi-logarithmic representation in **Figure 5(b)** further reveals a pronounced asymmetry in the transport characteristics, manifested as a clear shift of the minimum current away from zero bias. As quantified in **Figure 5(c)**, the extracted offset voltage ($V_0$) shifts monotonically toward more negative values (–0.73 V, –1.39 V, and –1.50 V) with increasing NaOH concentration, providing direct evidence for the emergence of an internal built-in potential. This internal field is attributed to defect-induced polarization and asymmetric charge redistribution within the modified lattice. The observed transport characteristics are consistent with trap-limited conduction, approaching a trap-filled limit regime at higher bias, where defect states dominate carrier transport[50,51].

These electrically inferred defect states are further corroborated by ultrafast spectroscopic measurements (discussed in a later section), which independently probe carrier relaxation pathways associated with NaOH-induced electronic states.

In stark contrast, **Figure 5(d)** shows that pristine 2D biotite exhibits no measurable current within the µA range, underscoring its intrinsically insulating nature. The emergence of a robust, bias-activated conduction upon NaOH treatment thus signifies a chemically driven transition from an insulating to a semiconducting state in a naturally abundant layered silicate, enabled by defect engineering and alkali-ion incorporation.

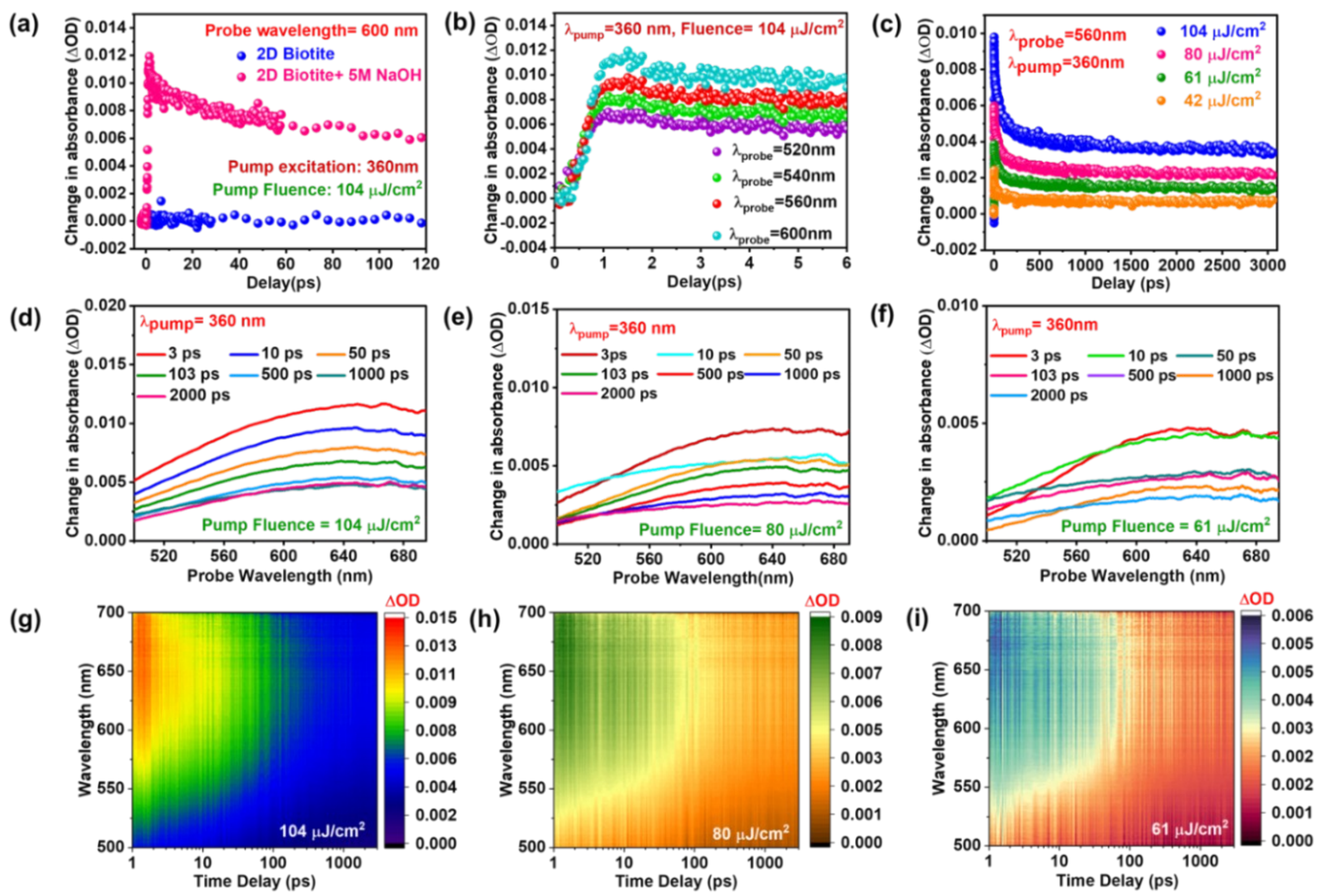

***Figure 6.*** *(a) Transient change in absorbance (ΔOD) as a function of pump–probe delay for pristine 2D biotite and 2D biotite treated with 5 M NaOH, measured with 360 nm pump excitation and a probe wavelength of 600nm at a pump fluence of 104 μJ/cm$^2$ (b) ΔOD dynamics recorded at a pump wavelength of 360 nm and constant pump fluence of 104 μJ/cm$^2$ for different probe wavelengths, (c) ΔOD as a function of delay time measured at a fixed pump wavelength of 360 nm and probe wavelength of 560 nm under different pump fluences (d-f) Transient absorption spectra (ΔOD vs probe wavelength) at selected delay times measured at different pump fluences;(g–i) Two-dimensional contour plots of ΔOD as a function of probe wavelength and pump–probe delay, acquired at a pump wavelength of 360 nm for different pump fluences.*

**Figure 6(a)** shows that pristine 2D biotite exhibits no detectable transient absorption (TA) signal under 360 nm (3.44 eV) excitation. Considering its wide electronic bandgap of ~5.2 eV, the pump photon energy is insufficient to drive direct interband transitions, which likely accounts for the absence of a measurable photoinduced response within the investigated spectral window. In contrast, 5 M NaOH-treated 2D biotite exhibits a pronounced ΔOD signal at a probe wavelength of 600 nm, demonstrating that alkali treatment substantially modifies the electronic structure of the layered biotite framework [52]. Previous characterization indicates that NaOH treatment reduces the bandgap to ~3.3 eV, consistent with a transition from the initially insulating state toward semiconducting behavior. Consequently, 360 nm excitation provides sufficient photon energy for above-bandgap excitation, enabling the generation of photoexcited carriers. The resulting nonequilibrium carrier population gives rise to a measurable transient absorption response, reflecting the excited-state dynamics of the chemically modified material. As shown in **Figure 6(b)**, the ΔOD dynamics are recorded at different probe wavelengths while maintaining a fixed pump wavelength of 360 nm and a constant pump fluence of 104 μJ $cm^{-2}$. **Figure 6(c)** presents the ΔOD dynamics measured at different pump fluences with the pump and probe wavelengths fixed at 360 and 560 nm, respectively.

**Figures 6(d)–(f)** show that the transient spectra exhibit a pronounced positive ΔOD response in the 500–700 nm region, consistent with excited-state absorption (ESA). Because the pump photon energy (3.44 eV) slightly exceeds the bandgap (~3.3 eV) of the treated sample, photoexcitation promotes carriers across the bandgap into conduction-band states. The resulting nonequilibrium carrier population can undergo further probe-photon absorption to higher-lying electronic states, giving rise to the observed ESA[53]. Notably, the persistence of the ESA response up to 2000 ps indicates that the photoinduced absorbing population is not completely depleted within the experimental time window, suggesting the involvement of long-lived photoexcited states and/or trapped carrier populations in the transient absorption response.

The two-dimensional TA contour maps (**Figures 6(g)–(i)**) further visualize the persistence of the positive ΔOD response across the probed spectral and temporal windows. Increasing the pump fluence enhances the signal amplitude without substantially altering its spectral or temporal characteristics, indicating that the dominant photoinduced relaxation pathways remain largely unchanged over the investigated excitation-density range. The approximately linear scaling of the ESA amplitude with pump fluence (**Figure S10**, Supporting Information) is consistent with a response proportional to the photoexcited carrier population and supports a contribution from free-carrier absorption (FCA) [54,55]. The long-lived component may additionally involve carrier trapping or defect-associated states introduced or modified during NaOH treatment.

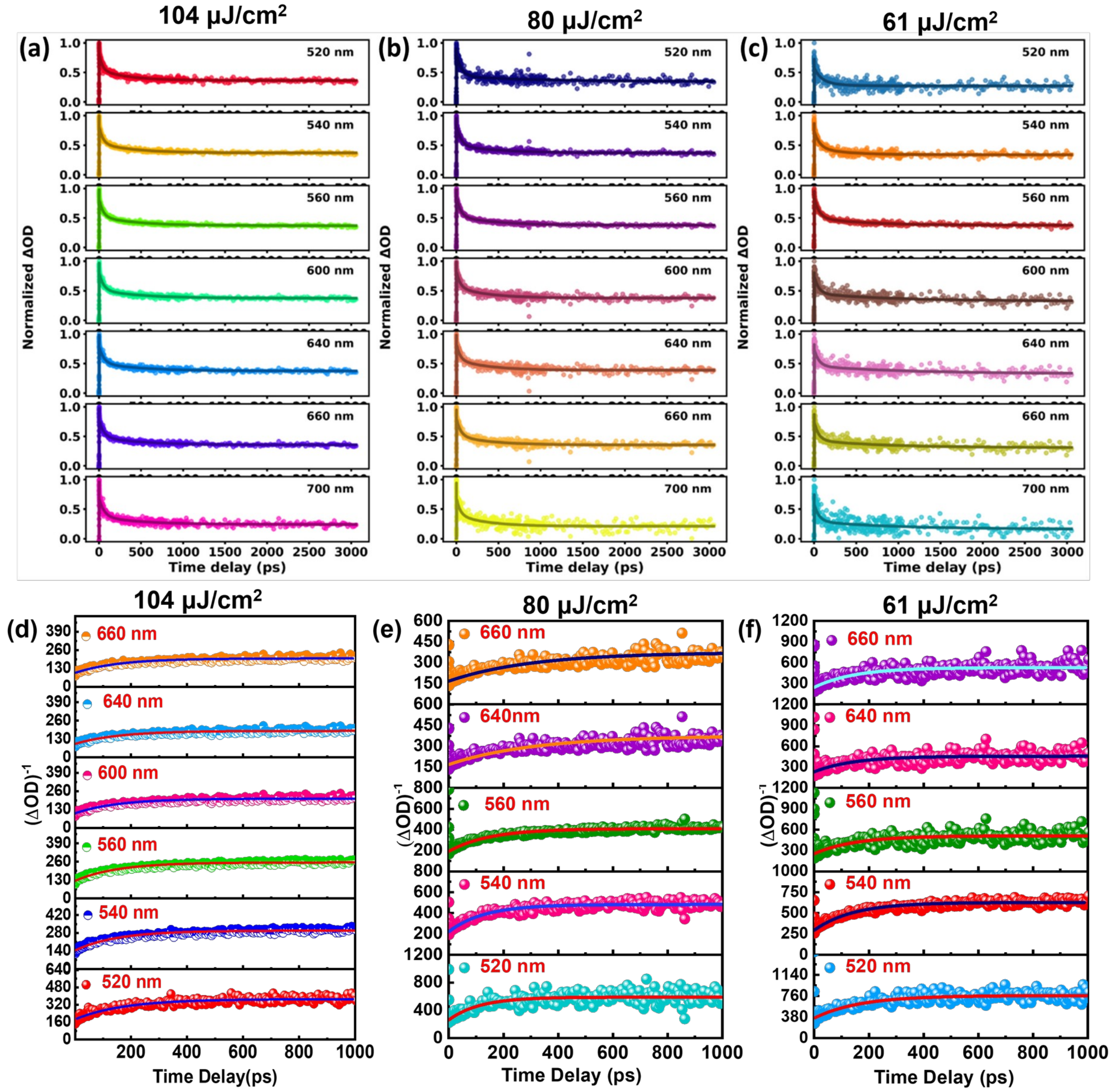

***Figure 7.*** *(a–c) Transient absorption kinetic decay traces of NaOH-treated 2D biotite recorded at different probe wavelengths under 360 nm excitation with pump fluences of 104, 80, and 61* μJ/cm$^2$, *respectively. (d–f) Corresponding temporal evolution of (ΔOD)$^{-1}$ at selected visible probe wavelengths with exponential fitting, measured under identical excitation conditions.*

**Figures 7(a–c)** show the transient kinetic traces of NaOH-treated 2D biotite recorded at different probe wavelengths under 360 nm excitation at pump fluences of 104, 80, and 61 μJ cm$^{-2}$, respectively. The kinetic traces at each probe wavelength were analyzed using an instrument-response-function (IRF)-convoluted model comprising an exponential rise followed by a biexponential decay, expressed as [56]

$\Delta A(t)=\left[\left(1-e^{-t/\tau_r}\right).\ H(t).\ \left(A_1 e^{-t/\tau_1}+A_2 e^{-t/\tau_2}\right)\right]\otimes IRF(t)$,where, IRF is the Gaussian instrument response function, $\tau_r$ is the rise time constant, H is a Heaviside function, $\tau_1$ and $\tau_2$ are the fast and slow decay time constants, respectively. The kinetic parameters obtained at different probe wavelengths under 360 nm pump excitation at a pump fluence of 104 μJ $cm^{-2}$ are summarized in **Table 1**, while the corresponding time constants for the other pump fluences are provided in **Table S2**.

The rise component reflects the ultrafast cooling of the photoexcited carriers following their rapid initial thermalization[57]. At a pump fluence of ~104 μJ $cm^{-2}$, the carrier cooling times extracted from the kinetic traces across the visible probe wavelengths are in the range of 0.16 –0.38 ps. The subsequent relaxation dynamics exhibit biexponential decay, with time constants of ~35–60 ps[58] and ~336–491ps[59]. The decay of the positive excited-state absorption (ESA) signal reflects the relaxation, trapping, and recombination of the photoexcited carriers. To elucidate the dominant recombination dynamics, the temporal evolution of $(\Delta OD)^{-1}$ was analyzed at different probe wavelengths (**Figure 7(d–f)**). The inverse transient signal exhibits an exponential increase with time delay, consistent with first-order carrier-recombination kinetics and suggesting a trap-mediated recombination pathway [60].[57]

This relaxation behavior is consistent with a defect-rich electronic structure arising from Na incorporation and the associated lattice perturbations. Accordingly, the fast decay component, $\tau_1$ (35–60 ps) is attributed to rapid carrier relaxation and trapping into shallow Na-induced localized states,[58,59]. In contrast, the slower component, $\tau_2$ (336–491ps) is attributed to longer-lived trap-assisted recombination involving deeper defect states introduced or modified by Na incorporation.[59]

**Time Constants vs Fluence**

The dependence of the characteristic time constants on pump fluence provides insight into the relaxation pathways governing the excited-state dynamics as presented in **Figure S11.** The rise time increases monotonically with increasing fluence at all probe wavelengths, consistent with a carrier-density-dependent slowing of carrier cooling. This behavior can be attributed to a hot-phonon bottleneck effect [61], in which the increased photoexcited carrier density enhances the generation and subsequent reabsorption of nonequilibrium phonons, thereby retarding carrier cooling.

In contrast, the fast decay component, $\tau_1$ exhibits only weak fluence dependence, with a slight decrease at shorter probe wavelengths (520–560 nm), while no systematic dependence is observed at longer probe wavelengths. The slow component $\tau_2$ similarly shows no clear dependence on excitation density across the investigated spectral range. likewise shows no clear dependence on excitation density across the investigated spectral range. The largely fluence-independent decay dynamics are consistent with predominantly monomolecular (first-order) carrier relaxation/recombination, suggesting that trap-mediated processes play a major role rather than higher-order carrier–carrier recombination mechanisms [62].

Such behavior is consistent with a defect-rich semiconducting system, in which localized states introduced or modified by chemical treatment can act as trapping and recombination centers. The combined fluence-dependent dynamics therefore suggest that, over the investigated excitation-density range, the photoexcited carriers in NaOH-treated 2D biotite undergo ultrafast cooling followed by predominantly trap-associated relaxation and recombination.

The dependence of the transient absorption amplitude on excitation density was further analyzed by plotting log($\Delta OD_{max}$) as a function of log(pump power) as presented in ***Figure S12***. The linear behavior confirms a power-law relationship, $\Delta OD_{max} \propto P_{pump}^{m}$, with extracted slopes of 0.64 (520 nm), 1.49 (560 nm), 1.34 (600 nm), 1.27 (640 nm), and 1.15 (700 nm). Slopes close to unity indicate a linear response governed by single-photon absorption[63].

**Table 1.** Kinetic parameters of NaOH-treated 2D biotite obtained by fitting the transient absorption traces at different probe wavelengths following 360 nm pump excitation, including the rise time ($\tau_r$), fast decay time ($\tau_1$), and slow decay time ($\tau_2$).

| Pump Fluence ($\mu J/cm^2$) | Probe wavelength (λ) (nm) | $\tau_r$ (ps) | $\tau_1$ (ps) | $\tau_2$ (ps) |
|---|---|---|---|---|
| 104 | 520 | 0.38 ± 0.06 | 35.66 ± 3.35 | 336.23 ± 37.96 |
| | 540 | 0.38 ± 0.04 | 43.05 ± 2.57 | 405.92 ± 36.35 |
| | 560 | 0.27 ± 0.02 | 46.37 ± 2.43 | 445.22 ± 38.34 |
| | 600 | 0.36 ± 0.02 | 51.25 ± 3.31 | 491.89 ± 59.88 |
| | 640 | 0.31 ± 0.03 | 54.89 ± 3.89 | 511.17 ± 70.79 |
| | 660 | 0.29 ± 0.02 | 53.06 ± 3.53 | 487.12 ± 60.85 |
| | 700 | 0.17 ± 0.07 | 60.20 ± 6.57 | 486.75 ± 52.45 |

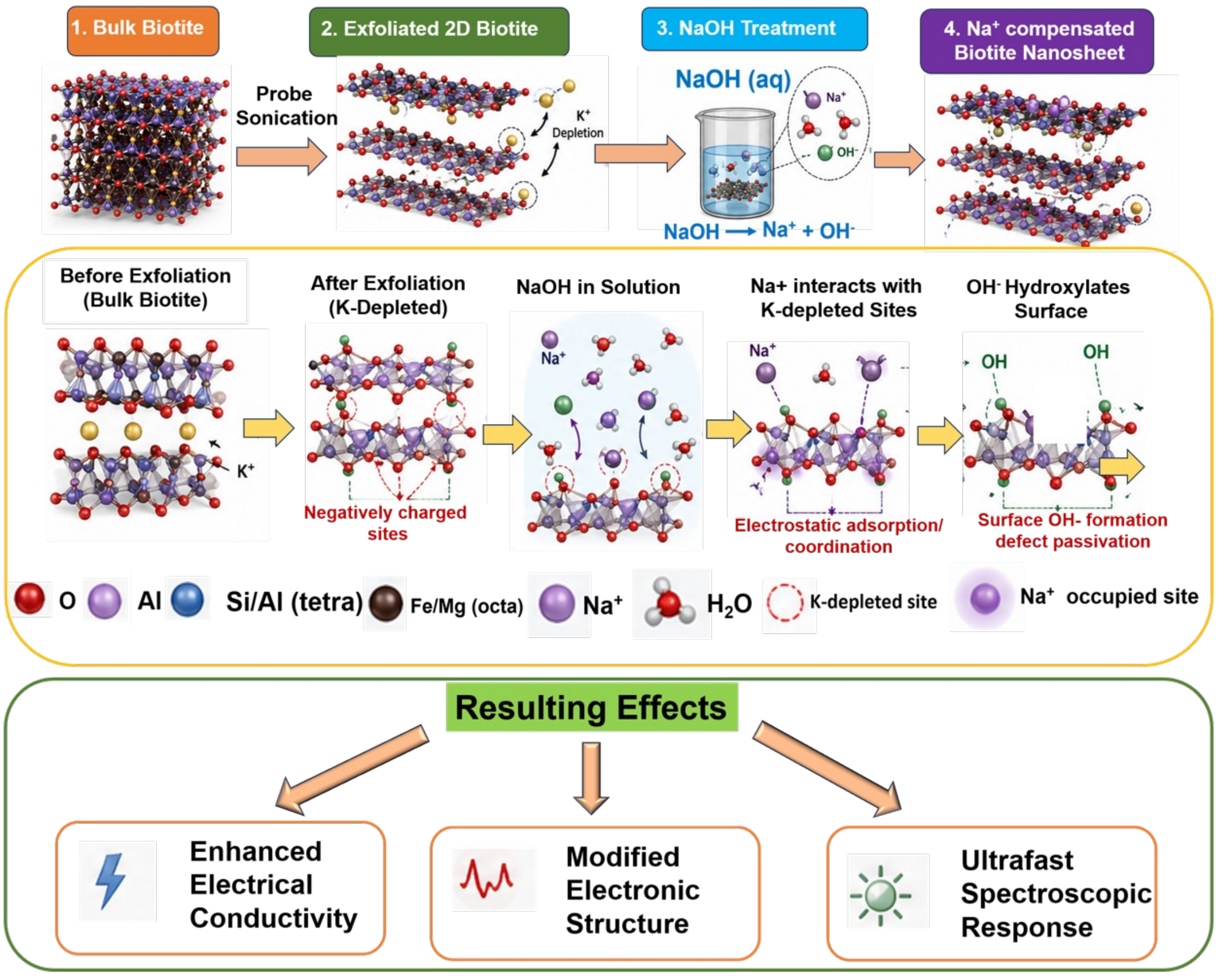

**Figure 8.** *Schematic of the NaOH-induced transformation mechanism in 2D biotite. $Na^+$ incorporation and partial $K^+$ exchange lead to interlayer contraction, while $OH^-$ induces surface bond reconfiguration and hydroxylation, resulting in a structurally modified Na-compensated biotite with semiconducting behavior confirmed by I–V and pump–probe measurements.*

The NaOH-induced transformation of 2D biotite is governed by coupled ion exchange, surface reconstruction, and interlayer reorganization processes. Upon exposure to NaOH solution, dissociated $Na^+$ and $OH^-$ ions interact with the exfoliated layered aluminosilicate framework, triggering simultaneous structural and electronic modifications.

At the interlayer region, partial exchange between interlayer $K^+$ ions and $Na^+$ occurs, driven by the comparable charge but smaller ionic radius of $Na^+$. This substitution reduces the effective size of interlayer cations and weakens the electrostatic separation between adjacent layers, resulting in a measurable contraction of the interlayer spacing and overall lattice shrinkage, consistent with XRD and DFT analysis.

Concurrently, $OH^-$ ions interact with surface metal–oxygen bonds (Si–O, Al–O, and Mg–O), promoting partial bond reconfiguration and enhanced surface hydroxylation. This leads to local distortion of the aluminosilicate framework and redistribution of charge density, as evidenced by systematic shifts in core-level binding energies in XPS spectra.

At the atomic scale, $Na^+$ ions preferentially occupy intrinsic pore sites within the 2D biotite lattice, where they are stabilized through strong electrostatic interactions with surrounding oxygen atoms. This site-selective occupation enables Na incorporation up to a critical concentration, beyond which additional Na species remain weakly adsorbed on the surface rather than being incorporated into the interlayer framework.

Importantly, the combined effect of Na incorporation and $OH^-$-induced structural reorganization leads to a significant modification of the electronic structure, resulting in the emergence of a Na-compensated biotite phase with semiconducting behavior. This transition is experimentally confirmed through I–V measurements, which show nonlinear current transport indicative of semiconducting characteristics, and is further supported by pump–probe spectroscopy, which reveals pronounced photoinduced carrier dynamics consistent with a finite bandgap and enhanced electronic response.

Overall, NaOH treatment transforms initially insulating 2D biotite into a structurally contracted, electronically redistributed, and semiconducting Na-compensated layered framework.

## Conclusion

In summary, we demonstrate that controlled NaOH treatment provides a simple and effective strategy for transforming naturally occurring biotite nanosheets from an insulating mineral into a chemically tunable 2D semiconductor. This insulator-to-semiconductor transition arises from the synergistic effects of Na incorporation, defect generation, and local structural reconstruction, while largely preserving the monoclinic layered framework. Complementary XRD, Raman, and XPS analyses reveal lattice contraction, interlayer reorganization, defect formation, hydroxylation, increased electron density, and partial $Na^+$–$K^+$ exchange, collectively establishing the structural and chemical origins of the electronic transformation.

The resulting chemical restructuring drives a pronounced evolution in the optical and electronic properties. The characteristic ~221 nm absorption of pristine 2D biotite evolves into broader features centered near ~280 and ~975 nm, extending the optical response from the UV toward the visible–NIR region. This evolution is accompanied by a substantial reduction in the optical bandgap from ~5.2 eV to ~3.2–3.5 eV and the emergence of low-energy transitions at ~1.12–1.17 eV, pointing to defect- and dopant-induced electronic states. Electrical measurements further demonstrate nonlinear conduction, microampere-level currents, reduced threshold voltages, and asymmetric transport in the treated nanosheets.

The electronic transformation is further manifested in the ultrafast optical response. Whereas pristine biotite exhibits negligible transient absorption, NaOH-treated nanosheets display pronounced excited-state absorption associated with the newly introduced electronic states. The excited-state dynamics encompass sub-picosecond carrier cooling (~0.16–0.38 ps), followed by fast (~35–60 ps) and long-lived (~336–491 ps) decay components associated with trap-mediated recombination. The systematic increase in rise time with excitation fluence further reveals a hot-phonon bottleneck that suppresses carrier cooling at elevated carrier densities.

Overall, this work establishes chemical defect and ion engineering as an effective route for converting naturally abundant layered minerals into functional 2D semiconductors with tunable electronic and ultrafast optical properties. The combined bandgap modulation, nonlinear charge transport, pronounced excited-state absorption, and prolonged carrier dynamics highlight NaOH-modified biotite as a versatile materials platform with potential for broadband photodetection, nonlinear and ultrafast optical modulation, photonic technologies, and hot-carrier applications.

## Experimental Section

**Synthesis Method:** A liquid-phase exfoliation technique is employed to derive 2D layers from bulk biotite. In this process, 15 mg of bulk biotite is dispersed in 60 mL of isopropyl alcohol (IPA) and sonicated at room temperature using a Rivotek probe sonicator for durations of 4h. After sonication, the samples are left undisturbed for 24h to allow the precipitate to settle at the bottom of the container. The supernatant containing the exfoliated layers is then separated via centrifugation and used for further analysis.

### NaOH Treatment of 2D Biotite

Exfoliated 2D biotite nanosheets were dispersed in isopropanol (IPA) via sonication to form a homogeneous suspension. Aqueous NaOH solutions (2 M, 5 M, and 10 M) were added to the dispersion, and the mixtures were maintained under continuous stirring at room temperature for 12 h. This process enables $Na^+$ interaction with the biotite structure through ion-exchange and surface adsorption mechanisms. The resulting dispersions were not subjected to any separation or thermal treatment.

**Characterization of Samples:** The XRD pattern of 2D biotite and NaOH treated samples are recorded using a Bruker D8 Advance X-ray diffractometer equipped with a LynxEye detector. Cu Kα radiation with a wavelength of 0.15406 nm is used, and the measurements are carried out over a 2θ range of 20-100°. Absorption spectra are measured using an Analytical UV–Vis 2080Plus double-beam spectrophotometer with quartz cuvettes of 10 mm path length. Room-temperature Raman spectra are recorded using a WITEC UHTS 300 VIS Raman spectrometer (Germany), operated with an excitation wavelength of 532 nm. The sample composition is analyzed using a PHI 5000 VersaProbe III scanning X-ray photoelectron spectroscopy (XPS) microprobe. A Jeol JSM-IT300HR is used to capture SEM images showing thin flakes of NaOH -treated biotite nanoflakes.

An HRTEM FEI Themis 60−300, coupled with an FEI-CETA 4k × 4k camera, is used to examine crystallographic plane orientations and surface defects. The thickness of the 2D flakes is determined using an Oxford Asylum Research MFP 3D Atomic Force Microscope at room temperature.

**I-V measurement:** A flexible polyethylene terephthalate (PET) substrate (20 × 10 mm) printed with Cu (12 µm)/Ni (1 µm)/Au (1 µm) interdigital electrodes (IDEs) was used for the device fabrication. The IDEs were patterned on a 75 µm-thick PET substrate (purchased from Newvision1981, Hong Kong, China) with a line width and inter-electrode gap of 100 µm and 30 interdigitated fingers. Exfoliated 2D biotite nanosheets dispersed in isopropanol (IPA) were deposited onto the IDEs by drop-casting and dried at room temperature. Similarly, NaOH-treated 2D biotite nanosheets were drop-cast onto the IDEs and dried under identical conditions. Current–voltage (*I–V*) measurements were performed using a Keithley 2480 source meter in a two-probe configuration.

**Transient Absorption Spectroscopy**: We perform transient absorption studies of NaOH treated 2D biotite using 360 nm pump excitation (generated through an OPA) and a broadband supercontinuum probe obtained by focusing the 808 nm Ti:Sapphire output into a sapphire crystal. The optical delay between the pump and probe pulses is tunable up to 8 ns. The time resolution of the setup is limited by the instrumental response function of ~150 fs, while the spectral resolution of the spectrometer is 0.8 nm. The pump and probe beams overlap at the sample position in a non-collinear geometry, with diameters of 2 mm and 100 µm, respectively. A chopper operating at 500 Hz is used to alternately allow and block the pump beam to obtain the differential absorption ($\Delta OD$) as:

$$\Delta OD(\lambda,\ t) = A^{pump}(\lambda,\ t) - A^{no\ \ pump}(\lambda)$$

Where $A^{pump}$ is the pump-induced absorption of the probe at a delay $t$ and wavelength$(\lambda)$, $A^{no\ \ pump}$ is the linear probe absorption. The efficiency of the setup is reported in our earlier work[64].The pump and probe beams are linearly polarized and have the same polarization.

## Computational Methods

To investigate the effects of sodium incorporation on the structural, electronic, and optical behavior of biotites, a series of first-principles simulations was performed. The computational approach aimed to elucidate how Na adsorption and substitution change the atomic arrangement, electronic band structure, and optical response of both bulk and monolayer configurations. All calculations were based on Density Functional Theory (DFT) with the Hubbard correction (DFT+U) as implemented in the SIESTA code[65,66] . The Perdew-Burke-Ernzerhof (PBE) functional within the Generalized Gradient Approximation (GGA) framework was employed, together with a double-ζ polarized (DZP) basis set. The real-space integration grid was defined with a mesh cutoff energy of 450 Ry, and the Brillouin zone (BZ) was sampled using a Γ-centered Monkhorst-Pack k-point mesh. Structural optimizations were carried out until the maximum residual force on each atom was below 0.05 eV/Å, and electronic self-consistency was achieved when the difference between successive density matrices was less than $10^{-4}$. For Fe-containing systems, a Hubbard parameter U of 4.0 eV was applied to correct for the localized *d*-electron Coulomb interactions.

For the monolayer configurations, additional *ab initio* simulations were performed using the same computational parameters and convergence thresholds described above. A 12×12×1 Monkhorst–Pack *k*-point mesh was employed to ensure adequate sampling of the Brillouin zone, and a vacuum buffer spacing of 30 Å was introduced along the z-axis to prevent spurious interactions between periodic (mirror) images. These conditions ensured reliable convergence of both the electronic structure and the optical response within the SIESTA framework.

To identify the most favorable adsorption sites and understand the interaction landscape of Na on the biotite surfaces, a systematic mapping of adsorption energies was performed. The Na atom was positioned at multiple in-plane coordinates covering the entire two-dimensional unit cell. In contrast, its position along the *z*-axis was fully optimized for each $(x,y)$ point until the atomic forces acting on Na were converged. The substrate atoms were kept fixed during these calculations to isolate the adsorbate-surface interaction. The adsorption energy at each point was obtained as the difference between the total energy of the Na-biotite system and the sum of the energies of the isolated components. The resulting energy is shown in adsorption maps that reveal the variation of the potential energy surface across the biotite monolayer.

The simulated Raman spectra for pristine biotite monolayers and for those with adsorbed Na atoms were calculated using the CRYSTAL23 package[67] . Vibrational frequencies at the Γ point were obtained from numerical second derivatives of the total energy. At the same time, Raman intensities were determined via the coupled perturbed Kohn-Sham (CPKS) method, which computes the polarizability tensors within a perturbative formalism[68].

For the optical properties analysis, the absorption coefficient (α) was calculated as a function of the photon energy (ω) using the following expression[69] :

$$\alpha(\omega)=\sqrt{2}\,\omega\cdot\left[\left(\epsilon_1^2(\omega)+\epsilon_2^2(\omega)\right)^{1/2}-\epsilon_1(\omega)\right]^{1/2},$$

where $\varepsilon_1(\omega)$ and $\varepsilon_2(\omega)$ are the real and imaginary components of the dielectric function, respectively. The dielectric response was normalized to exclude vacuum contributions in the out-of-plane direction, following recommended procedures for 2D materials[70] .

**Acknowledgements**

D.M. and P.K.D. thankfully acknowledge the UPM (SGDRI) project of IIT Kharagpur for all the equipments used in TAS measurements. C.S.T. acknowledges DAE Young Scientist. Research Award (DAEYSRA) and AOARD grant no. FA2386- 21-1-4014 and Naval Research Board for funding support. C.S.T. acknowledges the funding support of AMT and Energy& Water Technologies of TMD Division of DST. Guilherme S. L. Fabris acknowledges the São Paulo Research Foundation (FAPESP) fellowship (process number 2024/03413-9). Raphael B. de Oliveira thanks the National Council for Scientific and Technological Development (CNPq) (process number 200257/2025-0), Douglas S. Galvão acknowledges the Center for Computing in Engineering and Sciences at Unicamp for financial support through the FAPESP/CEPID Grant (process number 2013/08293-7). We thank the Coaraci Supercomputer Center for computer time (process number 2019/17874-0). Douglas S. Galvao also acknowledges support from INEO/CNPq and FAPESP grant 2025/27044-5. We thank the Coaraci Supercomputer Center for computer time (process number 2019/17874-0). Marcelo L. Pereira Junior acknowledges financial support from FAPDF (grant 00193-00001807/2023-16), CNPq (grants 444921/2024- 9 and 308222/2025-3), and CAPES (grant 88887.005164/2024-00).

**Supporting Information**

Supporting Information is available from the author.

**References**

[1] K. S. Novoselov, A. K. Geim, S. V. Morozov, D. Jiang, Y. Zhang, S. V. Dubonos, I. V. Grigorieva, A. A. Firsov, *Science (1979).* **2004**, *306*, 666.

[2] Q. H. Wang, K. Kalantar-Zadeh, A. Kis, J. N. Coleman, M. S. Strano, *Nat. Nanotechnol.* **2012**, *7*, 699.

[3] Y. Wang, C. Zhao, X. Gao, L. Zheng, J. Qian, X. Gao, J. Li, J. Tang, C. Tan, J. Wang, X. Zhu, J. Guo, Z. Liu, F. Ding, H. Peng, *Nat. Mater.* **2024**, *23*, 1495.

[4] L. Li, Y. Yu, G. J. Ye, Q. Ge, X. Ou, H. Wu, D. Feng, X. H. Chen, Y. Zhang, *Nat. Nanotechnol.* **2014**, *9*, 372.

[5] X. Wang, S. Lan, *Adv. Opt. Photonics* **2016**, *8*, 618.

[6] A. VahidMohammadi, J. Rosen, Y. Gogotsi, *Science (1979).* **2021**, *372*, DOI 10.1126/science.abf1581.

[7] E. Pomerantseva, Y. Gogotsi, *Nat. Energy* **2017**, *2*, 17089.

[8] D. Deng, K. S. Novoselov, Q. Fu, N. Zheng, Z. Tian, X. Bao, *Nat. Nanotechnol.* **2016**, *11*, 218.

[9] L. Qu, Y. Liu, J.-B. Baek, L. Dai, *ACS Nano* **2010**, *4*, 1321.

[10] A. Rasyotra, S. Ghosh, R. T. Nair, P. Venkatram, A. Chowdhury, M. S. Alam, J. M. Kumar, K. Mukhopadhyay, S. Das, *Nature Sensors* **2026**, *1*, 305.

[11] P. L. Mahapatra, R. Tromer, P. Pandey, G. Costin, B. Lahiri, K. Chattopadhyay, A. P. M., A. K. Roy, D. S. Galvao, P. Kumbhakar, C. S. Tiwary, *Small* **2022**, *18*, DOI 10.1002/smll.202201667.

[12] D. Mitra, G. S. L. Fabris, R. Benjamim De Oliveira, R. Sadhukhan, J. K. Sarkar, M. M. Ferrer, M. L. P. Junior, P. L. Mahapatra, D. K. Goswami, G. Costin, D. S. Galvão, C. S. Tiwary, P. K. Datta, *Adv. Opt. Mater.* **2026**, *14*, DOI 10.1002/adom.202502660.

[13] D. Mitra, C. Campos de Oliveira, A. Kartsev, R. Sadhukhan, J. K. Sarkar, A. A. Safronov, D. K. Goswami, G. Costin, P. Alves da Silva Autreto, C. S. Tiwary, P. K. Datta, *Nanoscale* **2026**, *18*, 5482.

[14] P. L. Mahapatra, A. K. Singh, R. Tromer, K. R., A. M., G. Costin, B. Lahiri, T. K. Kundu, P. M. Ajayan, E. I. Altman, D. S. Galvao, C. S. Tiwary, *J. Mater. Chem. C Mater.* **2023**, *11*, 2098.

[15] D. Mitra, Md. N. Hasan, C. C. Gowda, G. Costin, C. S. Tiwary, D. Karmakar, P. K. Datta, *ACS Appl. Nano Mater.* **2025**, *8*, 8187.

[16] M. Wang, H. Wang, Q. Zhang, D. Chen, S. Wang, D. Wang, X. Wu, W. Gao, *ACS Nano* **2024**, *18*, 25813.

[17] Y. Zhao, K. Xu, F. Pan, C. Zhou, F. Zhou, Y. Chai, *Adv. Funct. Mater.* **2017**, *27*, DOI 10.1002/adfm.201603484.

[18] X. Yang, J. Ni, *Phys. Rev. B* **2005**, *71*, 165438.

[19] Y. Chen, N. Li, L. Wang, L. Li, Z. Xu, H. Jiao, P. Liu, C. Zhu, H. Zai, M. Sun, W. Zou, S. Zhang, G. Xing, X. Liu, J. Wang, D. Li, B. Huang, Q. Chen, H. Zhou, *Nat. Commun.* **2019**, *10*, 1112.

[20] Y. Xie, H. Wang, G. Xu, J. Wang, H. Sheng, Z. Chen, Y. Ren, C. Sun, J. Wen, J. Wang, D. J. Miller, J. Lu, K. Amine, Z. Ma, *Adv. Energy Mater.* **2016**, *6*, DOI 10.1002/aenm.201601306.

[21] L. Seidl, N. Bucher, E. Chu, S. Hartung, S. Martens, O. Schneider, U. Stimming, *Energy Environ. Sci.* **2017**, *10*, 1631.

[22] Y. Gong, H. Yuan, C.-L. Wu, P. Tang, S.-Z. Yang, A. Yang, G. Li, B. Liu, J. van de Groep, M. L. Brongersma, M. F. Chisholm, S.-C. Zhang, W. Zhou, Y. Cui, *Nat. Nanotechnol.* **2018**, *13*, 294.

[23] A. R. , E. J. P. , M. G. , S. J. , & S. J. M. Gonzalez-Elipe, *The Journal of Physical Chemistry, 92(12), 3471-3476* **1988**.

[24] S. J. Kerber, J. J. Bruckner, K. Wozniak, S. Seal, S. Hardcastle, T. L. Barr, *Journal of Vacuum Science & Technology A: Vacuum, Surfaces, and Films* **1996**, *14*, 1314.

[25] C. Elmi, S. Guggenheim, R. Gieré, *Clays Clay Miner.* **2016**, *64*, 537.

[26] J. Zhou, J. Zhou, Z. Wan, Q. Qian, H. Ren, X. Yan, B. Zhou, A. Zhang, X. Pan, W. Fang, Y. Ping, Z. Sofer, Y. Huang, X. Duan, *Nature* **2025**, *643*, 683.

[27] J. A. Rotole, P. M. A. Sherwood, *Surface Science Spectra* **1998**, *5*, 11.

[28] S. Wannaparhun, S. Seal, *J. Mater. Chem.* **2003**, *13*, 323.

[29] R. Wang, P. Li, W. Zhou, Y. Li, K. Gao, Y. Ouyang, *Mater. Chem. Phys.* **2024**, *318*, 129224.

[30] K.-F. Berggren, B. E. Sernelius, *Phys. Rev. B* **1981**, *24*, 1971.

[31] H. Qiu, T. Xu, Z. Wang, W. Ren, H. Nan, Z. Ni, Q. Chen, S. Yuan, F. Miao, F. Song, G. Long, Y. Shi, L. Sun, J. Wang, X. Wang, *Nat. Commun.* **2013**, *4*, 2642.

[32] D. Bouilly, J. L. Janssen, J. Cabana, M. Côté, R. Martel, *ACS Nano* **2015**, *9*, 2626.

[33] M. Jain, J. R. Chelikowsky, S. G. Louie, *Phys. Rev. Lett.* **2011**, *107*, 216806.

[34] M. Ishii, Y. Yamashita, S. Watanabe, K. Ariga, J. Takeya, *Nature* **2023**, *622*, 285.

[35] M. Amani, D.-H. Lien, D. Kiriya, J. Xiao, A. Azcatl, J. Noh, S. R. Madhvapathy, R. Addou, S. KC, M. Dubey, K. Cho, R. M. Wallace, S.-C. Lee, J.-H. He, J. W. Ager, X. Zhang, E. Yablonovitch, A. Javey, *Science (1979).* **2015**, *350*, 1065.

[36] X. Guo, *Physical Chemistry Chemical Physics* **2014**, *16*, 20420.

[37] P. Dev, Y. Xue, P. Zhang, *Phys. Rev. Lett.* **2008**, *100*, 117204.

[38] G. E. Walrafen, R. T. W. Douglas, *J. Chem. Phys.* **2006**, *124*, DOI 10.1063/1.2121710.

[39] A. Tlili, D. C. Smith, J.-M. Beny, H. Boyer, *A Raman Microprobe Study of Natural Micas*, **1989**.

[40] A. Wang, J. Freeman, K. E. Kuebler, *Raman Spectroscopic Characterization of Phyllosilicates*, **n.d.**

[41] S. Osswald, V. N. Mochalin, M. Havel, G. Yushin, Y. Gogotsi, *Phys. Rev. B* **2009**, *80*, 075419.

[42] A. Eckmann, A. Felten, I. Verzhbitskiy, R. Davey, C. Casiraghi, *Phys. Rev. B* **2013**, *88*, 035426.

[43] X. Zhang, Z. Shao, X. Zhang, Y. He, J. Jie, *Advanced Materials* **2016**, *28*, 10409.

[44] A. M. Rao, P. C. Eklund, S. Bandow, A. Thess, R. E. Smalley, *Nature* **1997**, *388*, 257.

[45] M. Bruna, A. K. Ott, M. Ijäs, D. Yoon, U. Sassi, A. C. Ferrari, *ACS Nano* **2014**, *8*, 7432.

[46] S. Roy, A. G. Joshi, S. Chatterjee, A. K. Ghosh, *Nanoscale* **2018**, *10*, 10664.

[47] X. Song, Y. Li, M. Yin, W. Yi, W. Liu, J. Li, G. Xi, *Nano Lett.* **2024**, *24*, 11683.

[48] R. C. Haddon, A. F. Hebard, M. J. Rosseinsky, D. W. Murphy, S. J. Duclos, K. B. Lyons, B. Miller, J. M. Rosamilia, R. M. Fleming, A. R. Kortan, S. H. Glarum, A. V. Makhija, A. J. Muller, R. H. Eick, S. M. Zahurak, R. Tycko, G. Dabbagh, F. A. Thiel, *Nature* **1991**, *350*, 320.

[49] J. Park, S. Song, Y. Yang, S.-H. Kwon, E. Sim, Y. S. Kim, *J. Am. Chem. Soc.* **2017**, *139*, 10968.

[50] H. T. Nicolai, M. Kuik, G. A. H. Wetzelaer, B. de Boer, C. Campbell, C. Risko, J. L. Brédas, P. W. M. Blom, *Nat. Mater.* **2012**, *11*, 882.

[51] W. Shockley, W. T. Read, *Physical Review* **1952**, *87*, 835.

[52] S. Singla, P. Joshi, G. I. López-Morales, K. Watanabe, T. Taniguchi, C. E. Dreyer, B. Chakraborty, *Advanced Materials* **2025**, *37*, DOI 10.1002/adma.202502342.

[53] S. He, P. Jin, T. Lian, H. Zhu, *Nature Reviews Methods Primers* **2026**, *6*, 34.

[54] Y. Yang, D. P. Ostrowski, R. M. France, K. Zhu, J. van de Lagemaat, J. M. Luther, M. C. Beard, *Nat. Photonics* **2016**, *10*, 53.

[55] S. Prodhan, K. K. Chauhan, T. Singha, M. Karmakar, N. Maity, R. Nadarajan, P. Kumbhakar, C. S. Tiwary, A. K. Singh, M. M. Shaijumon, P. K. Datta, *Appl. Phys. Lett.* **2023**, *123*, DOI 10.1063/5.0156843.

[56] E. Lopriore, C. Louca, A. Genco, I. Landa, D. Erkensten, C. J. Sayers, S. Brem, R. Perea-Causin, K. Watanabe, T. Taniguchi, C. Gadermaier, E. Malic, G. Cerullo, S. D. Conte, A. Kis, *Nat. Commun.* **2025**, *16*, 10710.

[57] S. Prodhan, K. K. Chauhan, T. Singha, M. Karmakar, N. Maity, R. Nadarajan, P. Kumbhakar, C. S. Tiwary, A. K. Singh, M. M. Shaijumon, P. K. Datta, *Appl. Phys. Lett.* **2023**, *123*, DOI 10.1063/5.0156843.

[58] A. Treglia, A. Olivati, V. Romano, A. Iudica, G. M. Paternò, I. Poli, A. Petrozza, *Adv. Energy Mater.* **2025**, *15*, DOI 10.1002/aenm.202404905.

[59] H. Wang, C. Zhang, F. Rana, *Nano Lett.* **2015**, *15*, 339.

[60] S. Prodhan, K. K. Chauhan, T. Singha, M. Karmakar, N. Maity, R. Nadarajan, P. Kumbhakar, C. S. Tiwary, A. K. Singh, M. M. Shaijumon, P. K. Datta, *Appl. Phys. Lett.* **2023**, *123*, DOI 10.1063/5.0156843.

[61] Y. Yang, D. P. Ostrowski, R. M. France, K. Zhu, J. van de Lagemaat, J. M. Luther, M. C. Beard, *Nat. Photonics* **2016**, *10*, 53.

[62] K. Pagano, J. G. Kim, J. Luke, E. Tan, K. Stewart, I. V. Sazanovich, G. Karras, H. I. Gonev, A. V. Marsh, N. Y. Kim, S. Kwon, Y. Y. Kim, M. I. Alonso, B. Dörling, M. Campoy-Quiles, A. W. Parker, T. M. Clarke, Y.-H. Kim, J.-S. Kim, *Nat. Commun.* **2024**, *15*, 6153.

[63] S. Bhattacharya, A. Ghorai, S. Raval, M. Karmakar, A. Midya, S. K. Ray, P. K. Datta, *Carbon N. Y.* **2018**, *134*, 80.

[64] M. Karmakar, S. Bhattacharya, S. Mukherjee, B. Ghosh, R. K. Chowdhury, A. Agarwal, S. K. Ray, D. Chanda, P. K. Datta, *Phys. Rev. B* **2021**, *103*, 075437.

[65] J. M. Soler, E. Artacho, J. D. Gale, A. García, J. Junquera, P. Ordejón, D. Sánchez-Portal, *Journal of Physics: Condensed Matter* **2002**, *14*, 2745.

[66] A. García, N. Papior, A. Akhtar, E. Artacho, V. Blum, E. Bosoni, P. Brandimarte, M. Brandbyge, J. I. Cerdá, F. Corsetti, R. Cuadrado, V. Dikan, J. Ferrer, J. Gale, P. García-Fernández, V. M. García-Suárez, S. García, G. Huhs, S. Illera, R. Korytár, P. Koval, I. Lebedeva, L. Lin, P. López-Tarifa, S. G. Mayo, S. Mohr, P. Ordejón, A. Postnikov, Y. Pouillon, M. Pruneda, R. Robles, D. Sánchez-Portal, J. M. Soler, R. Ullah, V. W. Yu, J. Junquera, *J. Chem. Phys.* **2020**, *152*, DOI 10.1063/5.0005077.

[67] A. Erba, J. K. Desmarais, S. Casassa, B. Civalleri, L. Donà, I. J. Bush, B. Searle, L. Maschio, L. Edith-Daga, A. Cossard, C. Ribaldone, E. Ascrizzi, N. L. Marana, J.-P. Flament, B. Kirtman, *J. Chem. Theory Comput.* **2023**, *19*, 6891.

[68] M. Ferrero, M. Rérat, R. Orlando, R. Dovesi, *J. Chem. Phys.* **2008**, *128*, DOI 10.1063/1.2817596.

[69] R. M. Tromer, L. D. Machado, C. F. Woellner, D. S. Galvao, *Physica E Low. Dimens. Syst. Nanostruct.* **2021**, *129*, 114586.

[70] A. Chaves, J. G. Azadani, H. Alsalman, D. R. da Costa, R. Frisenda, A. J. Chaves, S. H. Song, Y. D. Kim, D. He, J. Zhou, A. Castellanos-Gomez, F. M. Peeters, Z. Liu, C. L. Hinkle, S.-H. Oh, P. D. Ye, S. J. Koester, Y. H. Lee, Ph. Avouris, X. Wang, T. Low, *NPJ 2D Mater. Appl.* **2020**, *4*, 29.

## Table of Contents Entry

We demonstrate the chemical conversion of naturally abundant biotite nanosheets from an insulating mineral into a tunable 2D semiconductor through controlled NaOH treatment. Coupled ion exchange, Na incorporation, defect generation, and structural reconstruction reduce the bandgap from ~5.2 to ~3.2–3.5 eV, enable nonlinear electrical transport, and modulate ultrafast carrier dynamics, establishing natural layered silicates as a promising platform for functional 2D materials.

## Ion-Engineered Insulator-to-Semiconductor Transition in Natural 2D Biotite

## Table of Contents Image

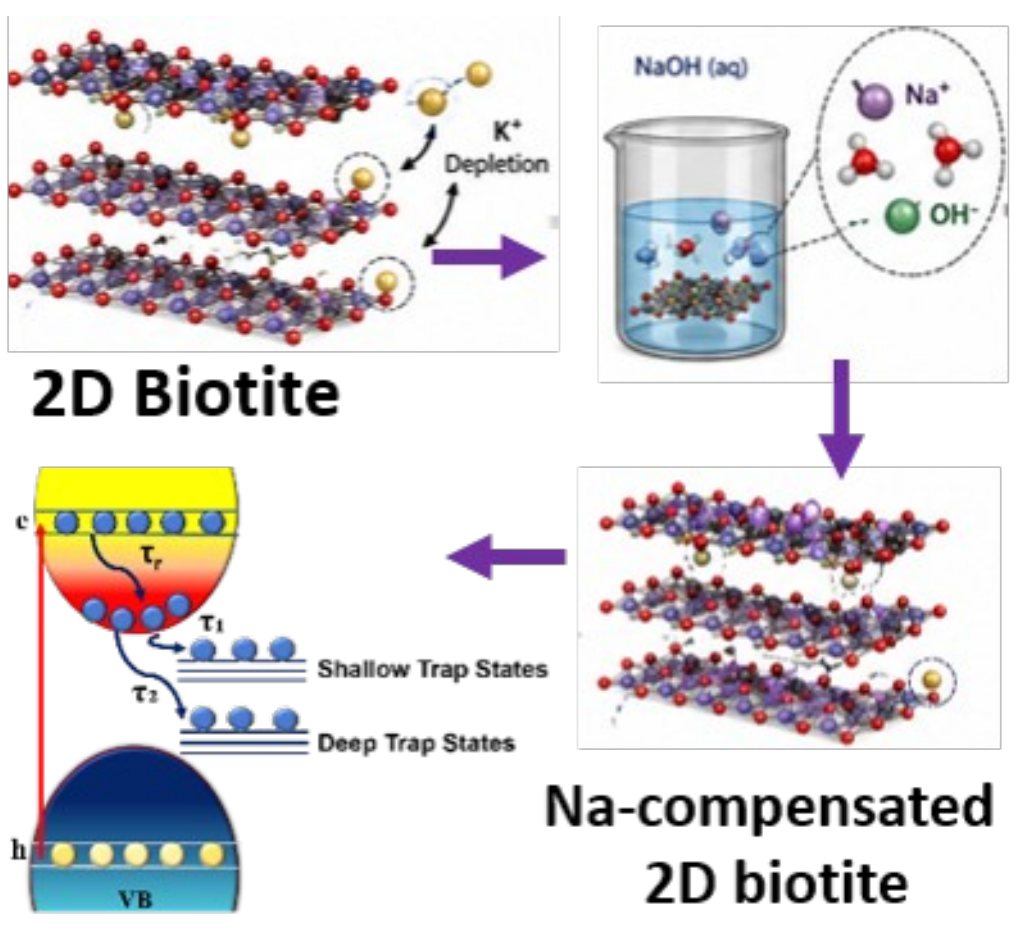